\documentclass[aps,prl,preprint,groupedaddress,showpacs]{revtex4}

\usepackage[utf8]{inputenc}
\usepackage{amsfonts,amsmath,amssymb}
\usepackage{graphicx}  % Add graphics capabilities
\usepackage{epstopdf} % to include .eps graphics files with pdfLaTeX
\usepackage{psfrag}
\usepackage[bookmarks,colorlinks=true,linkcolor=blue,urlcolor=black,citecolor=blue]{hyperref}

\begin{document}

%Title of paper
\title{Electric field noise above surfaces: a model for heating rate scaling law in ion traps}

\author{R. Dubessy}
\email[Electronic address: ]{romain.dubessy@polytechnique.edu}
\author{T. Coudreau}
\author{L. Guidoni}
%\homepage[]{Your web page}
%\thanks{}
%\altaffiliation{}
\affiliation{Laboratoire Mat\'{e}riaux et Ph\'{e}nom\`{e}nes Quantiques,\\
Universit\'{e} Paris Diderot et CNRS, UMR 7162,\\
10, rue Alice Domon et L\'{e}onie Duquet, 75013 Paris France}

%Collaboration name if desired (requires use of superscriptaddress
%option in \documentclass). \noaffiliation is required (may also be
%used with the \author command).
%\collaboration can be followed by \email, \homepage, \thanks as well.
%\collaboration{}
%\noaffiliation

\date{\today}

\begin{abstract}
We present a model for the scaling laws of the electric field noise spectral density as a function of the distance, $d$, above a conducting surface.
Our analytical approach models the patch potentials by introducing a correlation length, $\zeta$, of the electric potential on the surface.
The predicted scaling laws are in excellent agreement with two different classes of experiments (cold trapped ions and cantilevers), that span at least four orders of magnitude of $d$.
According to this model, heating rate in miniature ion traps could be greatly reduced by proper material engineering.
\end{abstract}

\pacs{37.10.Ty,34.35.+a,72.70.+m,37.10.Rs}
% insert suggested keywords - APS authors don't need to do this
%\keywords{}

%\maketitle must follow title, authors, abstract, \pacs, and \keywords
\maketitle

%\section{Introduction}
Device miniaturization is a challenge that raises new issues because the scaling laws valid in the macroscopic range might fail, for instance due to the emergence of a new characteristic length.
Even the simple case of the electric field in the vicinity of a conductor surface can exhibit anomalous behavior caused by small inhomogeneities of the electric potential on the surface.
These field fluctuations are crucial in the studies of short distance phenomena such as the measurement of the Casimir-Polder force\cite{Sukenik:1993qa}, studies of non contact friction\cite{Stipe:2001wc,Kuehn:2006kk}, gravitational forces\cite{Lockhart:1977mi} and contact potentials\cite{Camp:1991eg,Rossi:1992eq,Gaillard:2006uk}.

In a different context, recent success in quantum information experiment with trapped ions (see Ref~\cite{Blatt:2008vn} for a review) motivated the fabrication of micro-traps in order to fulfill the scalability requirement of a quantum computer\cite{DiVincenzo:2000uq}.
In such devices a set of micro-fabricated conducting electrodes generates an oscillating electric field that traps laser-cooled ions in a harmonic potential well, at a distance, $d$, of the surface.
In this situation the presence of a fluctuating electric field affects the ion motion inducing a heating, that is usually characterized in term of quanta of vibration gained per unit time.
This heating fixes a limit on the achievable fidelity of ion based quantum gates\cite{Wineland:1998}.
One might try to account for this heating rate by considering typical electric noise sources in conductors, among which the most likely is Johnson noise.
However, measured heating rates are orders of magnitude larger than the expected contribution of the Johnson noise.
Moreover, Johnson noise would induce a heating rate that scales as $d^{-2}$, whereas the observed one is consistent with a $d^{-4}$ scaling\cite{Leibfried:2003lq}, as would be expected from a random distribution of charges, leading to the notion of patch potentials\cite{Turchette:2000tx}.

Recent experiments\cite{Labaziewicz:2008on} suggest that indeed the surface quality plays a dominant role in this anomalous heating.
The observed scaling of the field noise with temperature points out thermally activated phenomena (surface defaults, charge traps,...).
These effects are easy to probe in the static limit where direct observation of the surface is possible using an atomic force microscope, but much more difficult to observe directly at higher frequencies, which are relevant to ion trapping.
However the measured noise can be compared by assuming a variation as $\omega^{-1}$, experimentally verified at room temperature\cite{Labaziewicz:2008on}.

The question of finding the electric field fluctuations near an infinite conductor filling half the space has already been extensively treated by considering thermal fluctuations carried by uncorrelated punctual sources (Johnson noise)\cite{Henkel:1999rr,Carminati:1999fv,Turchette:2000tx}.
In these models the characteristic length is given by the skin depth of the material, $\delta$, at the considered frequency.
The scaling of the electric field noise density is then expected to change from $(d/\delta)^{-3}$ for $d\ll\delta$ to $(d/\delta)^{-2}$ for $d\gg\delta$ where the typical value of $\delta$ for gold electrodes in the $MHz$ range is tens of microns.
However this behavior has not been observed in ion traps experiments where a $d^{-4}$ scaling is reported for $d$ varying from $75~\mu m$ to $1~mm$\cite{Epstein:2007fj}.
In a recent work\cite{Henkel:2008kl}, this type of model as been improved to take into account effects of charge diffusion, where the characteristic length, $\tilde{\delta}$, is related to the mean free path of charges on the surface. The predicted scaling is consistent with $(d/\tilde{\delta})^{-3}$ for $d\ll\tilde{\delta}$ to $(d/\tilde{\delta})^{-4}$ for $d\gg\tilde{\delta}$.

\paragraph{}
In this work we present a model where we take explicitly into account the spatial dependence of the electric field noise density above a surface, introducing a characteristic length not yet considered in previous models: the correlation length, $\zeta$, of the noisy potential on the surface.
In that picture, the noise arise from finite size sources distributed randomly on the surface.
We show that this simple model accounts for the behavior of electric field noise scaling on the whole range covered by both ion heating measurements and cantilever-based measurements. 
We propose that this correlation length could be related to the characteristic size of the patches, opening a way to the control of noise intensity by engineering the material properties.
In a recent work, an analytical solution of the Laplace equation for the special case of a planar ion trap was found\cite{House:2008sl}.
In the present letter we extend this approach and obtain an analytical expression for the scaling law of the electric field noise density $S_E(\omega,d)$ at a distance $d$ of the surface.
The first step is to solve the Laplace equation for the potential $\phi(x,y,z,t)$, $\Delta\phi(x,y,z,t)=0$ in half the space ($y>0$) with boundary condition $\phi(x,0,z,t)=\phi_0(x,z,t)$ and vanishing potential for $y\to\infty$. 
Under these assumptions, the potential reads:
\begin{eqnarray}
&&\nonumber\phi(x,y,z,t)=\int\frac{dk_x\,dk_z}{4\pi^2} e^{-y\sqrt{k_x^2+k_z^2}} \\
&&\times\int du\,dv \phi_0(u,v,t)e^{\imath(k_x(u-x)+k_z(v-z))}
\end{eqnarray}
where we took the Fourier transform of the Laplace equation, introducing $k_x$ and $k_z$.
Then carrying out the integration on $k_x$ and $k_z$, the resulting potential above the plane is:
\begin{equation}
\phi(x,y,z,t)=\int \frac{du\,dv}{2\pi} \phi_0(u,v,t)K(u-x,y,v-z)
\label{eqn:truepot}
\end{equation}
where we introduced $K(x,y,z)=\frac{y}{(x^2+y^2+z^2)^{3/2}}$, the kernel of the Laplace problem with these specific boundary conditions.
As in ref \cite{House:2008sl}, equation (\ref{eqn:truepot}) allows us to compute efficiently the potential associated to a given electrode geometry. 
More generally equation (\ref{eqn:truepot}) allows us to compute the potential created by any disordered boundary condition $\phi_0(x,z,t)$.

In the following we will assume that small patch potentials distributed over the plane $(x,z)$ create a disordered electric potential on the surface.
Let $\{C_i\}$ be the area of these patches and $V_i(t)$ the time-dependent electric noise on the patch $C_i$.
With these notations one have: $\phi_0(x,z,t)=\sum_{i} V_i(t) \chi_{C_i}(x,z)$ where $\chi_{C_i}$ is the characteristic function of $C_i$ ($\chi_{C_i}(x,z)=1$, $\forall(x,z)\in C_i$ and $\chi_{C_i}(x,z)=0$, $\forall(x,z)\notin C_i$).
From equation (\ref{eqn:truepot}) we can compute the electric field temporal correlation function at a distance $d$:
%\begin{widetext}
%\begin{equation}
\begin{eqnarray}
\nonumber
S_E(\tau,d)
&=&
\frac{1}{4\pi^2}\sum_i \sum_j\overline{V_i(t)V_j(t+\tau)}\\
\nonumber&&\times\int du^\prime dv^\prime \chi_{C_j}(u^\prime,v^\prime)\left[\nabla K\right](x-u^\prime,d,z-v^\prime)\\
&&\times\int dudv\chi_{C_i}(u,v)\left[\nabla K\right](x-u,d,z-v)
\label{eqn:corelfield}
\end{eqnarray}
%\end{equation}
%\end{widetext}
where $\left[\nabla K\right]$ is the gradient of the scalar field $K(x,y,z)$.
The horizontal line in equation (\ref{eqn:corelfield}) means that we average over many configurations of $V_i$.
We suppose that the noise on two distinct patches originates from independent random processes with the same temporal correlation function $R(\tau)$:
\begin{equation}
\overline{V_i(t)V_j(t+\tau)}=\delta_{i,j}R(\tau)
\end{equation}
Equation (\ref{eqn:corelfield}) can then be rewritten:
\begin{eqnarray}
\nonumber && S_E(\tau,d)=\frac{R(\tau)}{4\pi^2}\int dudv du^\prime dv^\prime \sum_i \chi_{C_i}(u,v)\chi_{C_i}(u^\prime,v^\prime)\\
&&\times\left[\nabla K\right](x-u,d,z-v).\left[\nabla K\right](x-u^\prime,d,z-v^\prime)
\label{eqn:corelTfield}
\end{eqnarray}
Since the set $\{C_i\}_i$ is disordered one can write: $\sum_i \chi_{C_i}(u,v)\chi_{C_i}(u^\prime,v^\prime)=N\left< \chi_{C_i}(u,v)\chi_{C_i}(u^\prime,v^\prime)\right>$, where the brakets denote an average on the configuration of $\{C_i\}$ and $N$ is the total number of patches.
Introducing the spatial correlation function of the patches, $C_\zeta(u-u^\prime,v-v^\prime)=\left< \chi_{C_i}(u,v)\chi_{C_i}(u^\prime,v^\prime)\right>$, where $\zeta$ is the (finite) correlation length, we take the temporal Fourier transform of $S_E(\tau,d)$:
\begin{eqnarray}
\nonumber &&
S_E(\omega,d)=\frac{NS_V(\omega)}{4\pi^2}\int \frac{dk_x dk_z}{4\pi^2} S_\zeta(k_x,k_z)\\
&&\times\left|\int dudv \left[\nabla K\right](x-u,d,z-v)e^{\imath(k_x u+k_z v)}\right|^2
\label{eqn:first}
\end{eqnarray}
where $S_V(\omega)$, defined as the Fourier transform of $R(\tau)$, is the potential noise spectral density on the surface and $S_\zeta(k_x,k_z)$ is the two dimensional Fourier transform of $C_\zeta(x,z)$.
After some calculations detailed in the Appendix, one obtains:
\begin{equation}
S_E(\omega,d)=\frac{NS_V(\omega)}{2\pi^2}\int dk d\theta S_\zeta(k\cos\theta,k\sin\theta)k^3e^{-2dk}
\label{eqn:final}
\end{equation}
To simplify further this equation we need an explicit form of $S_\zeta$.
In what follows we assume an exponential behavior for  the spatial autocorrelation function of the patches:
\begin{equation}
C_\zeta(x,z)=e^{-\sqrt{x^2+z^2}/\zeta}.
\label{eqn:distrib}
\end{equation}
This correlation function arises from a Poisson Voronoi tessellation model of the polycrystalline structure of metals\cite{Man:2006uq}.
More generally, following the approach of Ref \cite{Debye:1957rr}, it can be demonstrated that such an exponential behavior arises from a collection of random potential patches on a surface.

We can then compute the Fourier transform explicitly:
%\begin{equation}
$S_\zeta(k\cos\theta,k\sin\theta)=\frac{2\pi\zeta^2}{(1+\zeta^2k^2)^{3/2}}$,
%\end{equation}
finally giving:
\begin{equation}
S_E(\omega,d)=2\frac{N\zeta^2}{d} S_V(\omega)\int_0^\infty dk \frac{k^3 e^{-2k}}{(d^2+\zeta^2 k^2)^{3/2}}
\label{eqn:exprnoise}
\end{equation}

\begin{figure}[htbp]
   \centering
   \includegraphics[width=0.8\linewidth]{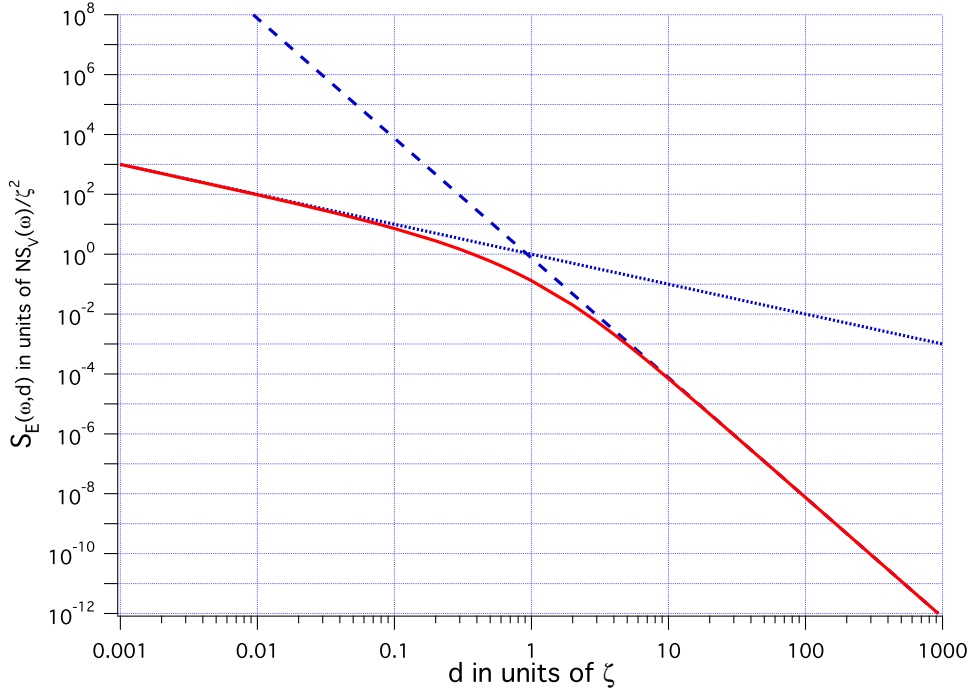} 
   \caption{(Color online) continuous line: normalized field noise density $S_E(\omega,d)$ (in units of $\frac{NS_V(\omega)}{\zeta^2}$) as $d$ varies for a fixed $\zeta$. For $d\ll\zeta$ one has $S_E(\omega,d)\sim d^{-1}$ (dashed line). For $d\gg\zeta$ one has $S_E(\omega,d)\sim d^{-4}$ (dotted line).}
   \label{fig:fixedZeta}
\end{figure}

%\section{Results}
Equation (\ref{eqn:exprnoise}) clearly identifies the separate contribution of the spatial and temporal components of the potential noise on the plane to the electric field noise density at a distance $d$ of the surface.
Let us point out two important limits: in the case $d\gg\zeta$, one finds $S_E(\omega,d)\approx \frac{3 \zeta^2 S_V(\omega)}{4 d^4}$ and in the case $d\ll\zeta$, one finds $S_E(\omega,d)\approx \frac{S_V(\omega)}{d\zeta}$.
Between these two simple asymptotic behaviors a smooth transition occurs as shown in Fig.~\ref{fig:fixedZeta}.
Let us note that the choice of any correlation function in (\ref{eqn:distrib}) would have led to the same long range limit ($d\gg\zeta$), whereas this choice is crucial for the short range behavior.

\paragraph{}
Let us now analyze how equation (\ref{eqn:exprnoise}) applies to both the case of ion traps and cantilevers based experiments.
In planar ion traps, charged particles are trapped above the surface in a tight pseudo-potential and are usually cooled down to the fundamental vibrational state.
The presence of a fluctuating electric field leads to the heating of the ion with a heating rate $\Gamma=\frac{e^2}{4m\hbar\omega}S_E(\omega,d)$, where $\omega$ is the vibrational frequency of the ion (typically a few $MHz$), $m$ is the mass, $e$ the charge and $d$ the equilibrium position of the ion\cite{Turchette:2000tx}.
As described in Refs\cite{Labaziewicz:2008on,Labaziewicz:2008ud}, the experimental measurement of ion heating rate gives thus access to the electric field noise density.
Due to electrode configuration in planar ion traps, the ions are trapped at a distance $d$ proportional to the size of the electrodes: the field can thus be probed in different ranges (typically $[75-150]~\mu m$) using different traps\cite{Labaziewicz:2008on}.
Although the surface quality (and thus $\zeta$) might depend on the fabrication process, a $d^{-4}$ dependence of the heating rate (or field noise density) has been observed.
These results are in agreement with the limit $d\gg\zeta$ of our model.

\paragraph{}
In cantilever based electric field noise measurement, the potential above a surface can be probed on a typical distance range from $10~nm$ to $400~nm$, yet on a very different scale from ion traps.
The cantilever is a device with resonant mechanical oscillation frequencies.
Its movement is damped by coupling to stray electric fields, with a rate $\Gamma=\frac{q^2}{4k_BT}S_E(\omega,d)$, where $\frac{\omega_c}{2\pi}$ is the frequency of the cantilever (typically a few $kHz$) and $q=CV$ is the induced charge, equal to the tip-sample capacitance, $C$, times the potential bias, $V$\cite{Kuehn:2006kk}.
Measuring the cantilever oscillations damping rate using optical interferometry gives access to the electric field noise density\cite{Stipe:2001wc,Kuehn:2006kk}.
These measurements, in the range $d\sim[10-100]~nm$, give a $d^{-1}$ scaling of the field noise density, consistent with the limit $d\ll\zeta$ of our model.

\paragraph{}
As mentioned above, the frequencies probed in ions traps and cantilever based measurement differ by several orders of magnitude, but the measured noise can be scaled to the same frequency, assuming a variation as $\omega^{-1}$.

Table~\ref{tab:booktabs} summarizes electric field noise density measurements above gold surfaces, as reported in several experimental works.
\begin{table}[htbp]
   \centering
   \begin{tabular}{@{} c|c|c|c|c @{}}
      \toprule
      Ref.  & $d$ & $\omega/2\pi$ & $S_E^{(exp)}(\omega,d)$ & $S_E(\omega_0,d)$ \\
      &$\mu m$&$MHz$&\multicolumn{2}{c}{$V^2m^{-2} Hz^{-1}$} \\\hline
      \cite{Stipe:2001wc} & $0.02$& $4\times10^{-3}$ & $4$ & $1.6\times10^{-2}$\\
      \hline
      \cite{Seidelin:2006io} & $40$ & $3$ & $9\times10^{-12}$ & $2.7\times10^{-11}$\\
      \cite{Labaziewicz:2008on} & $75$ & $1$ &  $[0.3-3]\times10^{-11}$ & $[0.3-3]\times10^{-11}$\\
      \cite{Turchette:2000tx} & $140$ & $10$ & $5\times10^{-12}$ & $5\times10^{-11}$\\
      \hline
   \end{tabular}
   \caption{Experimental room temperature $S_E(\omega,d)$ values taken from cited references. Above the horizontal line can be found a value measured using a cantilever. Below this line lie values measured in ion traps experiment. The last column give the rescaled value of $S_E(\omega_0,d)$ for $\omega_0/2\pi=1~MHz$. Data from reference \cite{Labaziewicz:2008on} have been obtained with the same trap after successive cleaning procedures and thermal cycling.}
   \label{tab:booktabs}
\end{table}

\begin{figure}[htbp]
   \centering
   \includegraphics[width=0.8\linewidth]{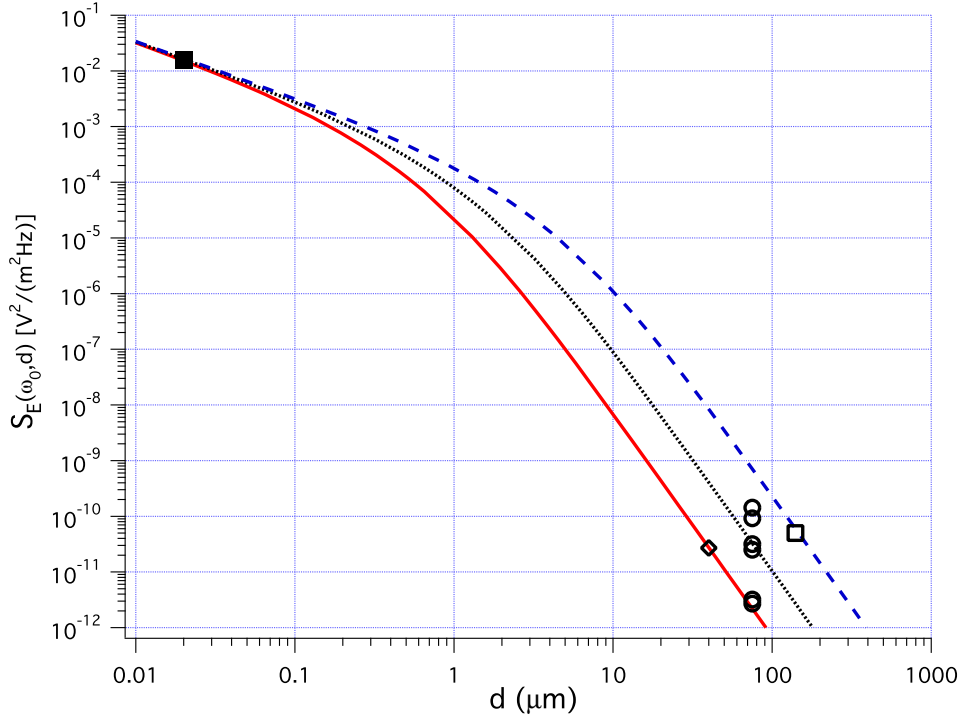} 
   \caption{Plot of experimental data from Refs.~\cite{Stipe:2001wc} ($\blacksquare$), \cite{Seidelin:2006io} ($\Diamond$), \cite{Labaziewicz:2008on} ($\bigcirc$) and \cite{Turchette:2000tx} ($\Box$). The lines correspond to the value of $S_E(\omega_0,d)$ predicted by our model for three values of $\zeta$, $\zeta=0.65\zeta_0$ (solid line), $\zeta=1.6\zeta_0$ (dashed line), $\zeta=4.6\zeta_0$ (dotted line).}
   \label{fig:modelVsdata}
\end{figure}

Let us compare our model to the rescaled measured values, $S_E(d)$, of Table~\ref{tab:booktabs} that we report in Fig.~\ref{fig:modelVsdata}.
In order to plot on the same graph the field noise density given by equation (\ref{eqn:exprnoise}), we need numerical values for $NS_V(\omega_0)$ and $\zeta$.
As the cantilever data are consistent with the short range limit of our model, $S_E(\omega_0,d)=NS_V(\omega_0)/(\zeta d)$ we can obtain a value for $NS_V(\omega_0)=3.2\times10^{-10}\zeta_0$, where we introduced $\zeta_0$ the characteristic length of this sample.
Under the simple assumption that $NS_V(\omega_0)$ is not sample dependent, we can plot the curves corresponding to different values of $\zeta/\zeta_0$ (dashed lines in Fig.~\ref{fig:modelVsdata}). 
We find that $0.6\zeta_0\leq\zeta\leq4.5\zeta_0$ covers the range of ion traps data.
As reported in Refs \cite{Camp:1991eg,Gaillard:2006uk}, the typical size of static patch potentials on gold surfaces is $\zeta_0=1~\mu m$ in excellent agreement with the assumption $\zeta_0\gg 20~nm$.
We interpret then the residual spread of the experimental data in terms of different correlation lengths (in the range $[0.6,4.5]~\mu m$) associated to the sample surface quality, itself highly dependent on the fabrication technique.
The role of surface quality has been noted in ref \cite{Labaziewicz:2008on}, where successive cleaning procedures decreased the measured electric field noise density (points $\bigcirc$ in Fig.~\ref{fig:modelVsdata}).
Thus our work shows that in order to compare measured field noise on very different scales, one has to take into account the correlation length of the material and not assume a $d^{-4}$ scaling of the noise, otherwise leading to an over-estimation of the noise at smaller distances.
For example, as shown in Fig.~\ref{fig:fixedZeta}, our model predicts that for a sample with $\zeta\sim1~\mu m$, the noise at $d=100~nm$ is $10^9$ times greater than noise at $d=100~\mu m$ (at the same frequency), three orders of magnitude lower than expected with a $d^{-4}$ scaling.

%\section{Conclusion}
\paragraph{}
In conclusion we developed a model for the electric field noise density based on an analytical approach, assuming a finite size of the potentials patches.
Electrical noise field density scaling laws predicted by this model are in very good agreement with experimental results, both in the long range regime (hundred of microns), probed with ion traps experiments and in the short range regime (tens of nanometers) probed with micro-fabricated cantilevers.
Let us point out that the pessimistic $d^{-4}$ scaling law for heating rates observed in trapped ion experiments is over estimated in the short range regime.
Moreover even though the noise considered in such devices lies within the $MHz$ range, we find a characteristic length compatible with measurement of static potential patches sizes.
This opens a new possibility to improve the performances of surface micro-traps based on the analysis of electric static noise obtained with atomic force microscopes.

% If you have acknowledgments, this puts in the proper section head.
\begin{acknowledgments}
We thank E. Boulat for fruitful discussions.
This work was supported by the French National Research Agency (ANR) Project No. ANR-JC0561454.
R.D. gratefully acknowledges the funding from the Délégation Générale de l'Armement (DGA).
\end{acknowledgments}

\appendix*
\section{Integrals}
Noticing that $K(d\times x,d\times y,d\times z)=d^{-2}K(x,y,z)$ we change variables in equation (\ref{eqn:first}) so that:
\begin{eqnarray*}
 &&
S_E(\omega,d)
=
\frac{NS_V(\omega)}{16\pi^4d^2}\int kdk d\theta S_\zeta(k\cos\theta,k\sin\theta)\\
&&\times
\left|\int udud\phi \left[\nabla K\right](u\cos\phi,1,u\sin\phi)\right.
\left.e^{-\imath dku\cos(\phi-\theta)}\right|^2
\end{eqnarray*}
where $u$ is now dimensionless.
Since:
\begin{equation*}
\left[\nabla K\right](u\cos\phi,1,u\sin\phi)=
\frac{1}{(1+u^2)^{5/2}}
\begin{pmatrix}
-3u\cos\phi \\
u^2 -2\\
-3u\sin\phi
\end{pmatrix},
\end{equation*}
$S_E(\omega,d)$ may now be rewritten:
\begin{eqnarray}
\nonumber &&
S_E(\omega,d)
=
\frac{NS_V(\omega)}{4\pi^2d^2}\int kdk d\theta S_\zeta(k\cos\theta,k\sin\theta)\\
&&\times
\left|\int_0^\infty du 
\frac{u}{(1+u^2)^{5/2}}
\begin{pmatrix}
-3\imath u\cos\theta J_1(dku) \\
(u^2-2)J_0(dku)\\
-3\imath u\sin\theta J_1(dku)
\end{pmatrix}
\right|^2
\end{eqnarray}
$J_n(x)$ being the $n$-th order Bessel function of the first kind.
The integral on $u$ can be analytically calculated leading to equation (\ref{eqn:final}).

% Create the reference section using BibTeX:

\end{document}